\documentclass[11pt]{article}

\usepackage{acl}
\usepackage{times}
\usepackage{latexsym}
\usepackage[T1]{fontenc}
\usepackage[utf8]{inputenc}
\usepackage{microtype}
\usepackage{inconsolata}
\usepackage{graphicx}
\usepackage{capt-of}
\usepackage{amsmath}
\usepackage{amssymb}
\usepackage{booktabs}
\usepackage{array}
\usepackage{multirow}
\usepackage{placeins}
\usepackage{afterpage}
\makeatletter
\def\section{\@startsection {section}{1}{\z@}{-1.8ex plus -0.5ex minus -.2ex}{1.2ex plus .3ex minus .2ex}{\large\bfseries\raggedright}}
\def\subsection{\@startsection{subsection}{2}{\z@}{-1.5ex plus -0.4ex minus -.2ex}{0.7ex plus .2ex minus .1ex}{\normalsize\bfseries\raggedright}}
\def\subsubsection{\@startsection{subsubsection}{3}{\z@}{-1.2ex plus -0.3ex minus -.2ex}{0.5ex plus .2ex minus .1ex}{\normalsize\bfseries\raggedright}}
\def\paragraph{\@startsection{paragraph}{4}{\z@}{1.5ex minus .2ex}{-1em}{\normalsize\bfseries}}
\def\subparagraph{\@startsection{subparagraph}{5}{\parindent}{1.5ex minus .2ex}{-1em}{\normalsize\bfseries}}
\makeatother
\title{BiMTokenizer: Preserving Semantic-Acoustic Balance in Low-Bitrate Speech Tokenization via Bidirectional State-Space Modeling}

\author{
  Xin Zhang$^{1}$ \quad
  Lin Li$^{1}$\thanks{Corresponding author.} \quad
  Chuanbo Liu$^{1}$ \quad
  Jianquan Liu$^{2,3}$ \quad
  Kong Aik Lee$^{4}$ \\
  $^{1}$Wuhan University of Technology, Wuhan, China \\
  $^{2}$NEC Laboratories Asia Pacific, Singapore
  \quad
  $^{3}$NEC Corporation, Japan \\
  $^{4}$The Hong Kong Polytechnic University, Hong Kong \\
  $^{1}$\texttt{\{zx361361,cathylilin,Lchb72\}@whut.edu.cn} \\
  $^{2}$\texttt{jianquan\_liu@nec.com.sg}
  \quad
  $^{3}$\texttt{jqliu@nec.com} \\
  $^{4}$\texttt{kong-aik.lee@polyu.edu.hk}
}

\begin{document}
\maketitle

\begin{abstract}
Speech codecs serve as bridges between continuous speech signals and large language models, yet face an inherent conflict between acoustic fidelity and semantic preservation. To mitigate this conflict, recent works increasingly adopt dual-tower architectures to decouple semantic and acoustic modeling with separate encoders. However, these dual-tower designs incur substantial architectural overhead. To avoid such complexity, we revisit the single-tower paradigm and propose \textbf{BiMTokenizer}, a low-bitrate speech codec (around 1.1 kbps) combining a \textbf{bidirectional state-space backbone} with \textbf{Residual Spherical Leech Quantization (RSLQ)}. The bidirectional backbone strengthens temporal modeling, while RSLQ offers a fixed, well-separated lattice bottleneck for robust semantic and acoustic tokenization without learned-codebook collapse. Experiments show that BiMTokenizer achieves superior acoustic reconstruction and the lowest WER among low-bitrate codec baselines across both clean and noisy environments, while using less than half the parameters of recent dual-tower baselines. Furthermore, its robust semantic representations yield strong performance on downstream speech understanding tasks, confirming that a well-designed single-tower codec can preserve the semantic--acoustic balance at low bitrates. The code and model weights are available at \url{https://github.com/ZhangXinWhut/BiMTokenizer}.
\end{abstract}

\section{Introduction}
\label{sec:introduction}

\begin{figure}[t]
  \centering
  \includegraphics[width=\linewidth]{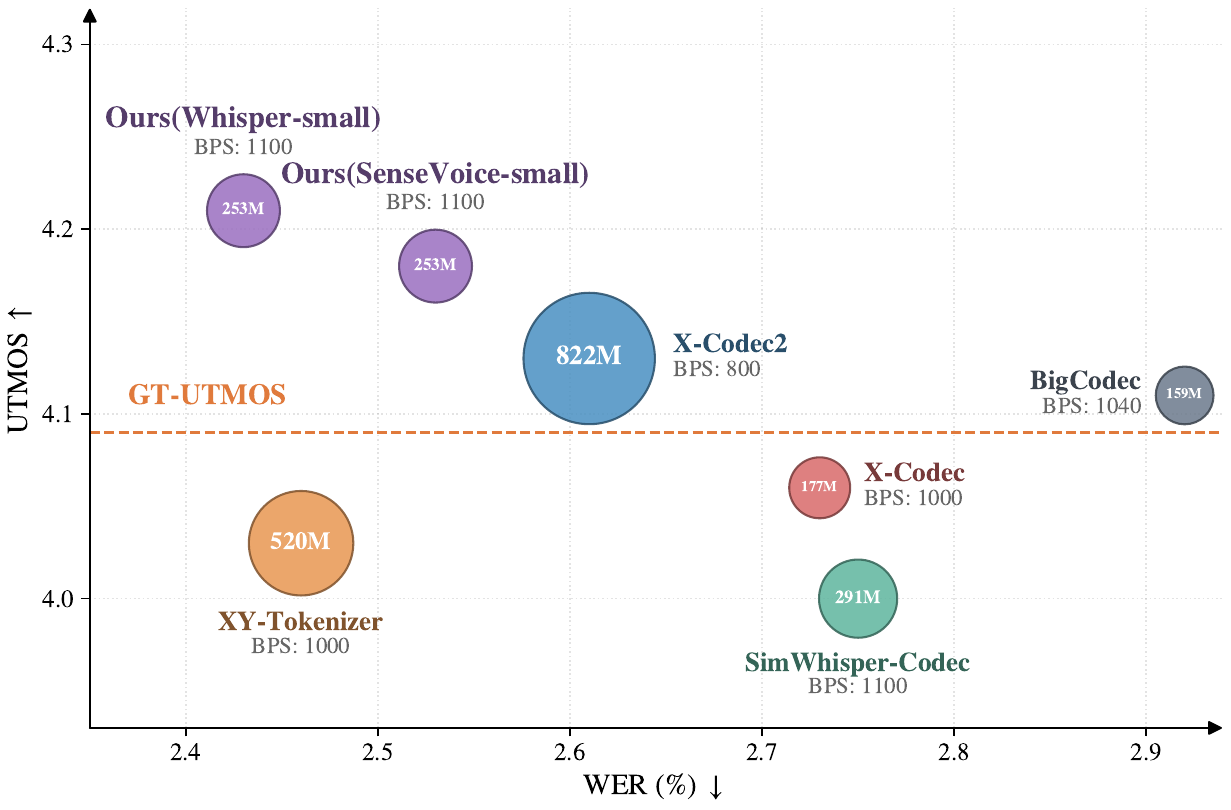}
  \caption{Codec comparison on speech reconstruction: WER (intelligibility, $\downarrow$) vs.\ UTMOS (naturalness, $\uparrow$); marker size indicates parameter count. Our variants are annotated by semantic teacher.}
  \label{fig:codec_comparison}
\end{figure}

In recent years, large language models (LLMs) have demonstrated remarkable performance in natural language processing, enabling fluent and natural text-based interactions~\citep{brown2020languagemodels}. Inspired by this success, researchers have extended LLMs to the speech modality, giving rise to Speech Large Language Models (Speech LLMs)~\citep{lakhotia2021gslm,borsos2023audiolm,wang2023valle,zhang2023speechgpt}. A key component of Speech LLMs is the speech codec, which converts continuous speech waveforms into discrete tokens compatible with token-based LLMs. Leveraging these discrete tokens, Speech LLMs perform autoregressive modeling over speech sequences, enabling a wide range of downstream speech tasks within a unified framework.

Existing speech tokens fall into two categories. Semantic tokens, derived from self-supervised learning (SSL) or supervised speech models~\citep{baevski2020wav2vec2,hsu2021hubert,chen2022wavlm,radford2023whisper,funaudiollm2024}, capture linguistic content but discard fine acoustic details. Acoustic tokens, in contrast, are produced by reconstruction-oriented neural audio codecs~\citep{zeghidour2021soundstream,defossez2022encodec,kumar2023dac}, preserving acoustic fidelity but showing weak semantic alignment with text-based LLMs. An ideal codec for Speech LLMs should support both, yet the two objectives tend to compete under low-bitrate budgets, a tension that has shaped much of the recent progress in speech codec design.

Early efforts stayed within the \textit{single-tower} paradigm: SpeechTokenizer~\citep{zhang2023speechtokenizer} distills SSL features into the first layer of a Residual Vector Quantization (RVQ) bottleneck, and Mimi~\citep{defossez2024moshi} adopts a split-RVQ structure that dedicates one codebook to semantic content and the rest to acoustic detail. Yet their effectiveness remains constrained at low bitrates: SpeechTokenizer was originally tuned for higher-bitrate regimes and degrades under aggressive compression, whereas Mimi, though designed natively for 1.1 kbps, still attains only a moderate semantic--acoustic balance in this regime. These limitations have shifted recent attention toward \textit{dual-tower} architectures~\citep{li2025dualcodec,chen2025sac}, which decouple semantic and acoustic modeling into separate encoders; the X-shape family~\citep{ye2025xcodec,ye2025llasa,gong2025xytokenizer} exemplifies this line, pairing a pretrained semantic encoder with an acoustic encoder and fusing their features before quantization. However, such dual-tower designs incur substantial architectural overhead, raising computational complexity relative to a single-tower codec. This raises a natural question: \textit{has the single-tower route truly been exhausted, or have its core components remained under-explored?} We pursue the latter, upgrading the two core components of the single-tower paradigm: the backbone and the quantizer.

In this work, we revisit these two components and propose \textbf{BiMTokenizer}, a low-bitrate ($\sim$1.1 kbps) single-tower speech codec built upon a bidirectional state-space backbone and Residual Spherical Leech Quantization (RSLQ). Speech evolves continuously along the time axis, and selective state-space models (SSMs) intrinsically capture such dynamics through structured temporal recurrence; equipping them with bidirectional scanning grants each frame access to both past and future context, providing the long-range temporal grounding required for faithful codec reconstruction. On the quantization side, RSLQ replaces the learnable codebooks of RVQ with a fixed, high-capacity codebook grounded in the spherical Leech lattice; its well-separated lattice geometry sidesteps codebook collapse by construction, while offering a representational space large enough to host semantic and acoustic information within a single token stream. Experiments show that BiMTokenizer attains superior acoustic reconstruction and the lowest WER among low-bitrate codec baselines under both clean and noisy conditions, while using fewer than half the parameters of recent dual-tower baselines. These results indicate that the semantic--acoustic balance in low-bitrate speech tokenization can be preserved not only through architectural separation, but also by strengthening the modeling and quantization design within a single-tower codec.

Our contributions can be summarized as follows:
\begin{itemize}
    \setlength\itemsep{0pt}
    \item We propose \textbf{BiMTokenizer}, a low-bitrate single-tower speech codec that couples a bidirectional state-space backbone with a fixed-lattice quantizer for joint semantic--acoustic tokenization.
    \item To the best of our knowledge, this work is the first to introduce Spherical Leech quantization into speech codecs to mitigate the semantic--acoustic conflict
    \item Experiments on LibriSpeech and the ARCH benchmark demonstrate that BiMTokenizer excels in both acoustic reconstruction and semantic retention. Further analysis on backbone design, quantizer choice, semantic supervision, and efficiency confirm the contribution of each proposed component.
\end{itemize}

\section{Related Work}
\label{sec:related}

\subsection{From Single-Tower to Dual-Tower Codecs}

To bridge continuous speech and LLMs, low-bitrate codecs must balance two competing objectives: preserving semantics for comprehension and maintaining acoustic fidelity for reconstruction. Early semantic-aware codecs mainly explored the single-tower route: SpeechTokenizer~\citep{zhang2023speechtokenizer} distills HuBERT~\citep{hsu2021hubert} features into a unified encoder--RVQ--decoder pipeline, while Mimi~\citep{defossez2024moshi} distills WavLM~\citep{chen2022wavlm} into the first level of a split RVQ, leaving the rest for acoustic residuals. SimWhisper-Codec~\citep{zhang2025simwhispercodec} repurposes a frozen, simplified Whisper encoder for low-bitrate speech coding without external semantic supervision. These designs keep inference simple, but their semantic--acoustic balance remains limited at low bitrates. Recent works therefore shift toward dual-tower (dual-encoder) codecs, including X-Codec/XCodec2~\citep{ye2025xcodec,ye2025llasa}, DualCodec~\citep{li2025dualcodec}, XY-Tokenizer~\citep{gong2025xytokenizer}, and SAC~\citep{chen2025sac}, which separate semantic and acoustic pathways via dedicated encoders or streams, at the cost of heavier architecture and optimization. BiMTokenizer questions whether this shift is inevitable. Rather than adding another tower, we strengthen the two shared components of the single-tower paradigm: the temporal backbone and the quantization bottleneck.

\subsection{Bidirectional State-Space Modeling in Speech}

Speech is inherently temporal, and codec reconstruction must preserve both long-range structure and local transients. State-space models (SSMs), from S4~\citep{gu2022s4} to Mamba~\citep{gu2023mamba}, provide a natural inductive bias via structured recurrent dynamics, whereas self-attention~\citep{vaswani2017attention} relies on content-based pairwise matching. In speech, bidirectional variants such as BiMamba have been used to replace or complement attention in ASR and enhancement~\citep{zhang2025mambaspeech,lin2026mambaexploration}, and Mamba-SEUNet exploits bidirectional dependencies for monaural enhancement~\citep{li2024mambaseunet}. Moreover, \citet{zhang2024rethinkingmamba} suggest that Mamba is particularly effective for reconstruction-oriented tasks (e.g., enhancement and spectrum reconstruction), while classification tasks may require additional modules. To our knowledge, no prior work uses bidirectional SSMs as the backbone of an end-to-end low-bitrate speech codec. BiMTokenizer fills this gap by using full-context bidirectional SSM blocks to improve reconstruction within a single stream.

\subsection{Quantization Methods for Discrete Tokenizers}

Modern speech codecs still rely largely on learnable Vector Quantization (VQ) and Residual Vector Quantization, inherited from Vector-Quantized Variational Autoencoders (VQ-VAEs)~\citep{vandenoord2017vqvae} and popularized by SoundStream~\citep{zeghidour2021soundstream}, EnCodec~\citep{defossez2022encodec}, and DAC~\citep{kumar2023dac}. However, learnable codebooks often require auxiliary objectives to maintain stable code utilization, and this issue becomes more pronounced when semantic and acoustic information are compressed into a single low-bitrate stream. Recent non-parametric visual tokenizers provide an alternative design space by replacing learned codebooks with fixed implicit ones. Lookup-Free Quantization (LFQ)~\citep{yu2024magvit2} and Binary Spherical Quantization (BSQ)~\citep{zhao2025bsq} remove learnable codebooks but still rely on entropy regularization for utilization; Finite Scalar Quantization (FSQ)~\citep{mentzer2024fsq} avoids both codebook learning and complex entropy penalties, yet its per-dimension level design remains heuristic and less natural for residual coding. Spherical Leech Quantization (SLQ)~\citep{zhao2026spherical} casts non-parametric quantization through the lens of lattice coding and uses the highly symmetric first shell of the Leech lattice to construct a large, fixed, and well-separated codebook, reducing the need for learned-codebook optimization and auxiliary codebook losses. Motivated by this geometry, BiMTokenizer adapts it to speech as RSLQ, providing a high-capacity fixed bottleneck for semantic supervision and acoustic residuals.

\begin{figure*}[!t]
  \centering
  \includegraphics[width=0.94\textwidth]{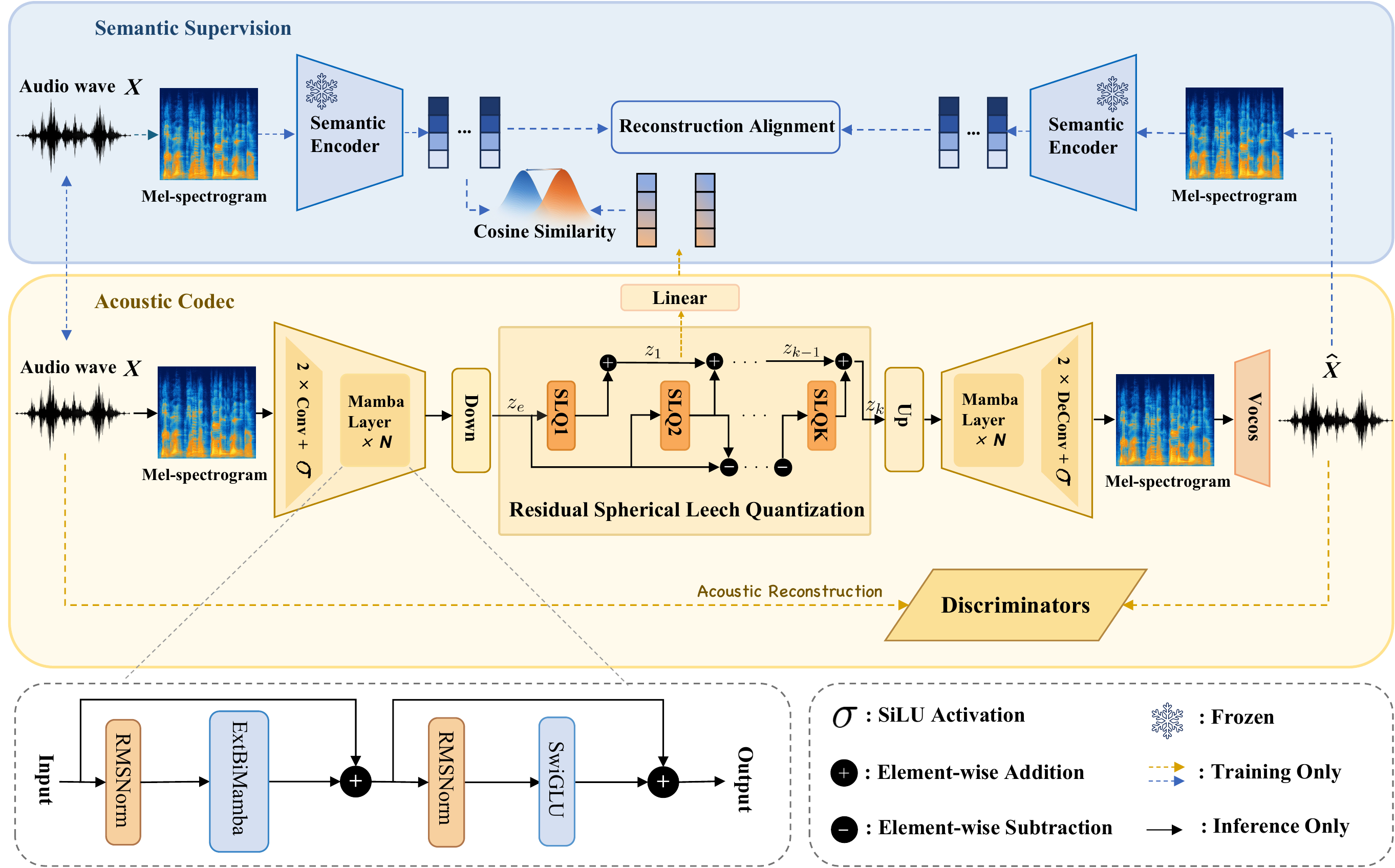}
  \caption{Overview of BiMTokenizer. The encoder and decoder use stacked bidirectional Mamba blocks, while the bottleneck is replaced by RSLQ. A frozen semantic teacher, instantiated as Whisper-small or SenseVoice-small, provides training-only semantic supervision and reconstruction alignment.}
  \label{fig:bimtokenizer}
\end{figure*}

\section{BiMTokenizer}
\label{sec:method}

To jointly support high-fidelity reconstruction and semantic preservation within a single-tower codec, we propose \textbf{BiMTokenizer}, which strengthens the two components where the semantic--acoustic conflict is most directly expressed: the shared temporal backbone and the shared quantization bottleneck. As illustrated in Fig.~\ref{fig:bimtokenizer}, BiMTokenizer is an end-to-end, single-stage, GAN-based neural speech codec that comprises a bidirectional Mamba encoder--decoder backbone, a Residual Spherical Leech Quantization (RSLQ) bottleneck, and an auxiliary semantic supervision branch driven by a frozen semantic encoder. The semantic branch is active only during training, while the acoustic codec operates autonomously at inference. We describe each component and the training objectives in the following subsections.

\subsection{Model Architecture}
\label{sec:architecture}

BiMTokenizer is a single-tower encoder--quantizer--decoder speech codec, as shown in Fig.~\ref{fig:bimtokenizer}. The encoder converts an input mel-spectrogram into a 50 Hz feature sequence through a two-layer convolutional stem and a stack of bidirectional Mamba blocks, after which a lightweight downsampler reduces the sequence to 12.5 Hz. At this rate, RSLQ encodes each frame into one semantic-aligned token followed by several residual acoustic tokens. The quantized representation is then upsampled and processed by a mirrored decoder of bidirectional Mamba blocks and transposed convolutions, producing a reconstructed mel-spectrogram that is converted to waveform by a Vocos vocoder~\citep{siuzdak2024vocos}. During training, a frozen semantic teacher (Whisper-small or SenseVoice-small)~\citep{radford2023whisper,funaudiollm2024} supervises the codec's representation via cosine alignment and a reconstruction-alignment loss; this branch is removed at inference, leaving a single 12.5 Hz discrete token stream. Further architectural details are provided in Appendix~\ref{sec:appendix-method}.

\subsection{Bidirectional Mamba Backbone}
\label{sec:bimamba}

A speech codec backbone must capture long-range temporal dependencies while preserving fine-grained local detail. Selective state-space models~\citep{gu2023mamba} provide a strong inductive bias for such sequential, locally-correlated signals at linear time complexity, and recent work shows that Mamba-based backbones can match or surpass attention-based counterparts on reconstruction-style speech tasks~\citep{zhang2024rethinkingmamba}. We therefore adopt Mamba as the temporal backbone of BiMTokenizer. Because we target the offline (non-streaming) codec setting, in which the entire utterance is available at encoding time, we adopt a \emph{bidirectional} variant so that each latent frame integrates context from both preceding and subsequent frames.

Let $h \in \mathbb{R}^{T \times d}$ denote the input hidden sequence to a backbone block. Each block is defined as
\begin{align}
  h' &= h + \mathrm{ExtBiMamba}(\mathrm{RMSNorm}(h)), \\
  h_{\mathrm{out}} &= h' + \mathrm{SwiGLU}(\mathrm{RMSNorm}(h')),
\end{align}
following the pre-norm, residual layout popularized by LLaMA~\citep{touvron2023llama}. Here $\mathrm{ExtBiMamba}(\cdot)$ denotes the external bidirectional Mamba layer of \citet{zhang2025mambaspeech}: the input and its time-reversed copy are processed by two Mamba branches with independent input and output projections, and the two outputs are fused by element-wise addition. Compared with vanilla unidirectional Mamba and the inner-fused InnBiMamba variant, ExtBiMamba has been reported to deliver stronger empirical performance on speech tasks while maintaining a simpler and faster realization~\citep{zhang2025mambaspeech}. The SwiGLU branch contributes per-frame non-linearity that complements the linear-in-time state-space operator, providing the additional expressive capacity required by the semantic-alignment objective in our single-tower design. We stack $N$ such blocks in both the encoder and the decoder.

\subsection{Residual Spherical Leech Quantization}
\label{sec:rslq}

To balance semantic preservation and acoustic reconstruction within a single low-bitrate stream, we propose \textbf{Residual Spherical Leech Quantization (RSLQ)}. RSLQ adapts the fixed spherical Leech codebook recently introduced for visual tokenization~\citep{zhao2026spherical} into a split residual bottleneck for speech codecs. Unlike standard RVQ, which learns its codebook embeddings, RSLQ quantizes against a fixed codebook $\mathcal{C}\subset\mathbb{S}^{23}$ on the 24-dimensional unit hypersphere, taken to be the $|\mathcal{C}|=196{,}560$ minimal vectors of the Leech lattice; this corresponds to $\log_2|\mathcal{C}|\approx 17.58$ bits per token. Since the Leech lattice realizes the densest known sphere packing in 24 dimensions~\citep{ConwayS88}, $\mathcal{C}$ provides a high-capacity yet uniformly distributed discrete space, which sidesteps the codebook-collapse issues commonly observed in learnable VQ at low bitrates~\citep{vandenoord2017vqvae}.

For each RSLQ layer, an input feature $f\in\mathbb{R}^{C}$ is projected to the 24-dimensional Leech space, normalized onto the unit hypersphere, quantized by maximum inner product, and mapped back to the codec feature space:
\begin{align}
\tilde{f}
&=
\frac{W_{\mathrm{down}}f}
{\|W_{\mathrm{down}}f\|_2}, \\
k
&=\arg\max_j \langle \tilde{f}, c_j\rangle,
\qquad c_j\in\mathcal{C},
\label{eq:rslq_quant}\\
e
&=\gamma W_{\mathrm{up}} c_k,
\end{align}
where $W_{\mathrm{down}}$ and $W_{\mathrm{up}}$ are learnable projections and $\gamma$ is a learnable scale. Gradients are propagated through the nearest-codeword assignment using a straight-through estimator~\citep{bengio2013estimating}. Since $\mathcal{C}$ is fixed, only the projection layers and scale parameters are learned.

RSLQ is designed around the semantic--acoustic structure of speech rather than used as a direct visual-tokenizer replacement. Given an encoder frame $u$, we allocate the first quantization layer to semantic modeling and the remaining layers to acoustic residual coding. The semantic layer quantizes $u$ in parallel and is supervised by the semantic distillation objective in Section~\ref{sec:semantic-supervision}. The acoustic layers form a residual chain:
\begin{align}
e_0&=\mathrm{RSLQ}_0(u),\qquad r_0=u,\\
e_i&=\mathrm{RSLQ}_i(r_{i-1}),\\
r_i&=r_{i-1}-e_i,\quad i=1,\dots,K-1,\\
\hat{u}&=e_0+\sum_{i=1}^{K-1}e_i.
\end{align}
This split residual design lets a single encoder assign one discrete layer to linguistic content while using the remaining layers for speaker traits, prosody, and fine acoustic detail, avoiding the architectural cost of a separate semantic tower. With a token rate of 12.5 Hz and $K=5$, BiMTokenizer operates at approximately $5\times12.5\times17.58 \approx 1.10~\mathrm{kbps}$.
Additional validation-time codebook statistics are provided in Appendix~\ref{app:rslq}.

\subsection{Auxiliary Semantic Supervision}
\label{sec:semantic-supervision}

To inject linguistic information through training-only supervision, BiMTokenizer uses a frozen semantic teacher $S_{\tau}(\cdot)$ only during training, where $\tau$ denotes either Whisper-small or SenseVoice-small. Given the input and reconstructed waveforms, we extract semantic targets
\begin{align}
  s &= S_{\tau}(x), & \hat{s} &= S_{\tau}(\hat{x}).
\end{align}
The semantic RSLQ output is mapped to the teacher feature space by a lightweight projector $P_{\psi}$~\citep{han2020coordinated,li2021coopnet}:
\begin{equation}
  \tilde{s}=P_{\psi}(q_{\mathrm{sem}}).
\end{equation}
Teacher-specific projector details are provided in Appendix~\ref{sec:appendix-semantic-projector}.
We use a frame-level cosine semantic distillation loss
\begin{equation}
  \mathcal{L}_{\mathrm{sem}}=
  1-\frac{1}{T}\sum_{t=1}^{T}
  \frac{\tilde{s}_t^{\top}s_t}{\|\tilde{s}_t\|_2\|s_t\|_2},
\end{equation}
Inspired by self-supervised representation reconstruction (SSRR)~\citep{lee2026ssrr}, we further enforce a reconstruction alignment loss on the synthesized audio $\hat{x}$ using a frozen ASR model as the feature extractor. This objective encourages the decoder output, rather than only the quantized bottleneck, to preserve the semantic information of the original speech:
\begin{equation}
  \mathcal{L}_{\mathrm{align}}=
  \frac{1}{T}\sum_{t=1}^{T}\|\hat{s}_t-s_t\|_1.
\end{equation}
Both semantic objectives are training-only. The frozen teacher, the projector, and the alignment path are discarded at inference.

\begin{table*}[t]
  \centering
  \scriptsize
  \setlength{\tabcolsep}{1.8pt}
  \resizebox{\textwidth}{!}{
  \begin{tabular}{lccccccccccccc}
    \toprule
    \multirow{2}{*}{Model} & Codebook & \multirow{2}{*}{bps} &
    Frame & \multirow{2}{*}{$N_q$} &
    Single & \multirow{2}{*}{Param} &
    \multirow{2}{*}{SIM$\uparrow$} & \multirow{2}{*}{STOI$\uparrow$} &
    PESQ & PESQ &
    \multirow{2}{*}{UTMOS$\uparrow$} & \multirow{2}{*}{ViSQOL$\uparrow$} &
    \multirow{2}{*}{WER{(\%)}$\downarrow$} \\
    & Size & & Rate & & Tower & & & &
    NB$\uparrow$ & WB$\uparrow$ & & & \\
    \midrule
    Ground Truth & -- & -- & -- & -- & -- & -- & 1.00 & 1.00 & 4.55 & 4.64 & 4.09 & 5.00 & 2.16 \\
    \midrule
    EnCodec (24 kHz) & 1024 & 1500 & 75 & 2 & \checkmark & 15M & 0.60 & 0.85 & 1.95 & 1.56 & 1.58 & 3.59 & 5.63 \\
    DAC (16 kHz) & 1024 & 1500 & 50 & 3 & \checkmark & 74M & 0.47 & 0.80 & 1.61 & 1.25 & 1.48 & 3.38 & 7.80 \\
    SpeechTokenizer (16 kHz) & 1024 & 1000 & 50 & 2 & \checkmark & 103.7M & 0.36 & 0.77 & 1.59 & 1.25 & 2.28 & 3.15 & 4.49 \\
    BigCodec (16 kHz) & 8192 & 1040 & 80 & 1 & \checkmark & 159M & 0.84 & \underline{0.94} & 3.27 & 2.68 & 4.11 & 4.21 & 2.92 \\
    Mimi (24 kHz) & 2048 & 1100 & 12.5 & 8 & \checkmark & 79M & 0.74 & 0.91 & 2.80 & 2.25 & 3.63 & 3.85 & 3.28 \\
    WavTokenizer (24 kHz) & 4096 & 900 & 75 & 1 & \checkmark & 80M & 0.65 & 0.90 & 2.63 & 2.13 & 3.79 & 3.74 & 4.18 \\
    SimWhisper-Codec (16 kHz) & 2016 & 1100 & 12.5 & 8 & \checkmark & 291M & 0.83 & 0.93 & 3.29 & 2.72 & 4.00 & 4.31 & 2.75 \\
    XCodec (16 kHz) & 1024 & 1000 & 50 & 2 & $\times$ & 177M & 0.68 & 0.86 & 2.68 & 2.11 & 4.06 & 3.72 & 2.73 \\
    XCodec2.0 (16 kHz) & 65536 & 800 & 50 & 1 & $\times$ & 820M & 0.82 & 0.92 & 3.04 & 2.43 & 4.13 & 3.83 & 2.61 \\
    DualCodec (24 kHz) & 16384/4096 & 1075 & 12.5 & 1/6 & $\times$ & 664M & 0.84 & 0.93 & 3.25 & 2.71 & 4.12 & 4.19 & \underline{2.46} \\
    XY-Tokenizer (16 kHz) & 1024 & 1000 & 12.5 & 8 & $\times$ & 520M & \underline{0.85} & 0.92 & 3.10 & 2.50 & 4.03 & 4.22 & \underline{2.46} \\
    \midrule
    BiMTokenizer-Whisper (Ours, 16 kHz) & 196560 & 1100 & 12.5 & 5 & \checkmark & 253M &
    \textbf{0.87} & \textbf{0.95} & \textbf{3.56} & \textbf{3.03} & \textbf{4.21} & \underline{4.33} & \textbf{2.44} \\
    BiMTokenizer-SenseVoice (Ours, 16 kHz) & 196560 & 1100 & 12.5 & 5 & \checkmark & 253M &
    \underline{0.85} & \underline{0.94} & \underline{3.45} & \underline{2.85} & \underline{4.18} & \textbf{4.34} & 2.53 \\
    \bottomrule
  \end{tabular}}
  \caption{Speech reconstruction comparison on LibriSpeech \textit{test-clean}. BiMTokenizer-Whisper and BiMTokenizer-SenseVoice use the same single-tower codec architecture and differ only in the frozen semantic teacher used during training. Best codec results are in bold, and second-best results are underlined.}
  \label{tab:test-clean}
\end{table*}

\begin{table*}[t]
  \centering
  \scriptsize
  \setlength{\tabcolsep}{5pt}
  \resizebox{\textwidth}{!}{
  \begin{tabular}{l|lccccccc}
    \toprule
    Category & Model & Token Rate & BPS &
    RAVDESS$\uparrow$ & EMOVO$\uparrow$ &
    SLURP$\uparrow$ & AM$\uparrow$ & Avg.$\uparrow$ \\
    \midrule
    \multirow{4}{*}{\textit{SSL Models}$^\dagger$}
    & wav2vec 2.0 & -- & -- & 55.32 & 31.80 & 14.37 & 86.38 & 46.97 \\
    & data2vec & -- & -- & 48.03 & 27.27 & \textbf{43.57} & 99.06 & 54.48 \\
    & HuBERT & -- & -- & \underline{65.28} & \underline{40.48} & \underline{33.75} & \textbf{99.58} & \underline{59.77} \\
    & WavLM & -- & -- & \textbf{67.94} & \textbf{43.08} & 30.98 & \underline{99.50} & \textbf{60.38} \\
    \midrule
    \multirow{2}{*}{\textit{ASR Models}}
    & Whisper-small & -- & -- & \textbf{83.33} & \underline{53.06} & \underline{63.30} & \underline{99.83} & \underline{74.88} \\
    & SenseVoice-small & -- & -- & \underline{82.64} & \textbf{56.29} & \textbf{71.47} & \textbf{99.98} & \textbf{77.60} \\
    \midrule
    \multirow{9}{*}{\textit{Codec Models}}
    & EnCodec & 150 & 1500 & 30.90 & \underline{27.72} & 8.37 & 72.57 & 34.89 \\
    & DAC & 150 & 1500 & 31.25 & 21.60 & 7.98 & 69.46 & 32.57 \\
    & BigCodec & 80 & 1040 & 31.94 & 15.99 & 7.72 & 65.83 & 30.37 \\
    & XCodec2.0 & 50 & 800 & 32.99 & 27.21 & 7.66 & 66.14 & 33.50 \\
    & XY-Tokenizer & 100 & 1000 & 39.58 & 24.49 & \underline{18.39} & \underline{96.34} & \underline{44.70} \\
    & WavTokenizer & 75 & 900 & 32.55 & \textbf{31.63} & 8.02 & 69.57 & 35.44 \\
    & SimWhisper-Codec & 100 & 1100 & \underline{42.71} & 24.15 & 8.13 & 79.76 & 38.69 \\
    & BiMTokenizer-Whisper & 62.5 & 1100 & \textbf{43.19} & 27.08 & 12.51 & 86.20 & 42.25 \\
    & BiMTokenizer-SenseVoice & 62.5 & 1100 & 40.28 & 26.87 & \textbf{18.48} & \textbf{98.02} & \textbf{45.91} \\
    \bottomrule
  \end{tabular}}
  \caption{Semantic representation evaluation on the speech domain of ARCH. Best results in each category are in bold, and second-best results are underlined. ASR models provide teacher upper-bound references; codec rows use pooled quantized representations. $^\dagger$ SSL results are taken from the original ARCH benchmark.}
  \label{tab:arch}
\end{table*}

\subsection{Training Objectives}
\label{sec:training-objectives}

BiMTokenizer is optimized with reconstruction, adversarial, feature-matching, and semantic objectives. We use a multi-scale mel-spectrogram reconstruction loss, together with a multi-period discriminator (MPD)~\citep{kong2020hifigan} and a multi-scale short-time Fourier transform discriminator (MS-STFTD)~\citep{defossez2022encodec}. The adversarial objective follows the least-squares generative adversarial network (GAN) formulation~\citep{mao2017lsgan}, and feature matching is computed on discriminator intermediate features. The generator objective is
\begin{equation}
\begin{split}
  \mathcal{L}_{G} =
  &\lambda_{\mathrm{rec}}\mathcal{L}_{\mathrm{rec}}
  + \lambda_{\mathrm{adv}}\mathcal{L}_{\mathrm{adv}}
  + \lambda_{\mathrm{feat}}\mathcal{L}_{\mathrm{feat}} \\
  &+ \lambda_{\mathrm{sem}}\mathcal{L}_{\mathrm{sem}}
  + \lambda_{\mathrm{align}}\mathcal{L}_{\mathrm{align}}.
\end{split}
\end{equation}
Detailed reconstruction, adversarial, and feature-matching losses are provided in Appendix~\ref{sec:appendix-objective}. Because RSLQ uses a fixed geometric codebook, the objective contains no VQ codebook loss, commitment loss, or entropy regularization.

\section{Experimental Setup}
\label{sec:experiments}

\subsection{Dataset and Training Details}
\label{sec:dataset}

Following common practice in low-bitrate neural speech codecs~\citep{xin2024bigcodec, defossez2022encodec}, we train BiMTokenizer on the full LibriSpeech~\citep{panayotov2015librispeech} training set (960 hours, sampled at 16\,kHz), using randomly cropped 2-second segments as input. The model contains 253M parameters in total. BiMTokenizer is instantiated with two choices of semantic teacher, yielding the variants BiMTokenizer-Whisper and BiMTokenizer-SenseVoice, which use Whisper-small~\citep{radford2023whisper} and SenseVoice-small~\citep{funaudiollm2024}, respectively, as the frozen teacher throughout training.

Training runs for 1{,}000{,}000 steps in a single stage on two NVIDIA H100 GPUs, with a per-GPU batch size of 64 and gradient accumulation of 1, giving an effective batch size of 128. Both the generator and the discriminator are optimized with AdamW~\citep{loshchilov2019adamw} ($\beta_1{=}0.8$, $\beta_2{=}0.9$, weight decay $0.01$), under a cosine-annealing schedule~\citep{loshchilov2017sgdr} that decays the learning rate from $1{\times}10^{-4}$ to $0$ after 30k warmup steps.  Additional training details are provided in Appendix~\ref{sec:appendix-objective}.

\subsection{Evaluation Details}
\label{sec:evaluation-details}

An effective speech codec for Speech LLMs should reconstruct speech with high fidelity while preserving discrete representations that remain useful for speech understanding. We therefore evaluate BiMTokenizer from two complementary perspectives: speech reconstruction and semantic representation.

\paragraph{Speech Reconstruction.}
We evaluate reconstruction quality on LibriSpeech \textit{test-clean}~\citep{panayotov2015librispeech}, reporting codec configuration, intelligibility, perceptual quality, and speaker similarity metrics: STOI, WER, PESQ-NB/PESQ-WB, UTMOS, ViSQOL, and SIM. Metric definitions, evaluation models, and implementation links are provided in Appendix~\ref{sec:appendix-evaluation}. The comparison includes representative low-bitrate speech codecs with similar operating ranges, covering reconstruction-oriented codecs~\citep{defossez2022encodec,kumar2023dac,xin2024bigcodec,defossez2024moshi}, semantic-aware single-tower tokenizers~\citep{zhang2023speechtokenizer,ji2024wavtokenizer,zhang2025simwhispercodec}, and dual- or X-shaped tokenizers~\citep{ye2025xcodec,ye2025llasa,li2025dualcodec,gong2025xytokenizer}.

\paragraph{Speech Representation.}
We evaluate the semantic quality of codec representations using the speech-domain protocol of the ARCH benchmark~\citep{laquatra2024arch}, following recent codec evaluations~\citep{ji2024wavtokenizer,chen2025sac}. ARCH covers emotion recognition with RAVDESS~\citep{livingstone2018ravdess} and EMOVO~\citep{costantini2014emovo}, intent classification with SLURP~\citep{bastianelli2020slurp}, and digit recognition with AudioMNIST~\citep{becker2024audiomnist}. For each codec, quantized representations are average-pooled over time and used as input to a linear classifier. To contextualize the gap between discrete codec tokens and continuous speech representations, we also report self-supervised learning (SSL) references, including wav2vec 2.0~\citep{baevski2020wav2vec2}, data2vec~\citep{baevski2022data2vec}, HuBERT~\citep{hsu2021hubert}, and WavLM~\citep{chen2022wavlm}. Since BiMTokenizer uses frozen ASR encoders as semantic teachers during training, we additionally include Whisper-small and SenseVoice-small as teacher upper-bound references rather than codec baselines.

\begin{figure*}[!t]
  \centering
  \includegraphics[width=\textwidth,height=0.21\textheight,keepaspectratio]{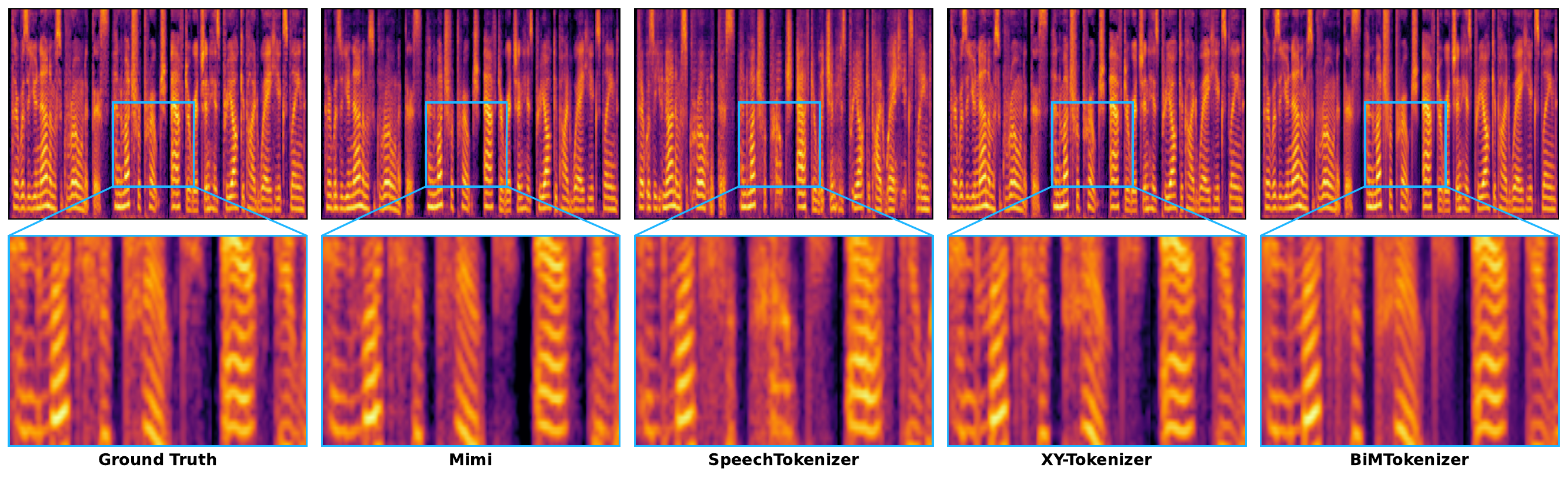}
  \caption{Mel-spectrogram reconstructions. The zoomed regions show that BiMTokenizer preserves local harmonic and spectral structure more faithfully than prior low-bitrate codecs.}
  \label{fig:spectrogram}
\end{figure*}

\section{Experimental Results and Discussion}
\label{sec:results}

\subsection{Speech Reconstruction Results}
\label{sec:reconstruction-results}

\paragraph{Quantitative Results.}

Table~\ref{tab:test-clean} compares BiMTokenizer with representative low-bitrate neural codecs on LibriSpeech \textit{test-clean}. At 1.1 kbps and 12.5 Hz, BiMTokenizer-Whisper achieves state-of-the-art reconstruction among compared low-bitrate codecs, with the best SIM, STOI, PESQ-NB, PESQ-WB, UTMOS, and WER. In particular, it reaches 3.56 PESQ-NB and 3.03 PESQ-WB, substantially exceeding all codecs below 1.5 kbps. BiMTokenizer-SenseVoice also obtains the best ViSQOL and remains close to the Whisper variant on most reconstruction metrics. Figure~\ref{fig:spectrogram} further shows that Mimi and SpeechTokenizer recover the dominant energy regions but lose some narrow harmonic bands, while XY-Tokenizer preserves more structure yet still exhibits visible smoothing of local transitions in the enlarged region. BiMTokenizer retains sharper bands and clearer transitions, making its reconstruction visually closer to the ground truth. This qualitative evidence complements the PESQ and UTMOS gains. Appendix~\ref{sec:appendix-robustness} reports consistent gains on the noisier LibriSpeech \textit{test-other} set and the out-of-distribution Seed-TTS-Eval benchmark in English and Mandarin~\citep{anastassiou2024seedtts}.

\paragraph{Variant Comparison.}

The two variants differ only in the frozen teacher. Whisper-small's 50 Hz features match the encoder's pre-downsampling rate; this denser supervision yields the strongest PESQ and WER. SenseVoice-small's lower-rate features align more closely with the 12.5 Hz bottleneck, delivering stronger semantic transfer, as shown in Table~\ref{tab:arch}. This contrast suggests that the teacher's frame rate, rather than its training objective alone, shapes how supervision interacts with the codec bottleneck.

\subsection{Semantic Representation Results}
\label{sec:arch-results}

Table~\ref{tab:arch} evaluates the semantic representation quality of codec tokens on the speech-domain tasks of ARCH. BiMTokenizer-SenseVoice obtains the highest average accuracy, surpassing the second-best XY-Tokenizer by 1.21 absolute points while remaining a single-tower codec. Its advantage is most clear on text-related tasks: it achieves the best codec results on SLURP and AudioMNIST, indicating that the quantized representations retain useful linguistic information even under a 1.1\,kbps bitrate. BiMTokenizer-Whisper, in turn, achieves the best codec result on RAVDESS, consistent with its stronger reconstruction in Table~1 and suggesting that denser Whisper supervision benefits paralinguistic cues. More broadly, both variants improve substantially over reconstruction-oriented codecs without explicit semantic supervision (EnCodec, DAC, BigCodec), confirming that the single-tower design yields competitive codec-level semantic representations on this benchmark.

\subsection{LLM-Based Speech Generation}
\label{sec:tts-generation}

We evaluate BiMTokenizer on Seed-TTS-Eval~\citep{anastassiou2024seedtts} using an autoregressive decoder-only model. Because the original tokenizer targets codec evaluation, we retrain a TTS-oriented variant for speech generation, as detailed in Appendix~\ref{sec:appendix-tts}. We report word error rate (WER), speaker similarity (SIM), and UTMOS on the English and Chinese test sets.

As shown in Table~\ref{tab:tts-generation}, BiMTokenizer achieves 1.87\% WER, 0.59 SIM, and 4.11 UTMOS on \textit{test-en}, together with 1.72\% WER, 0.65 SIM, and 3.30 UTMOS on \textit{test-zh}. These results confirm that its discrete representations support autoregressive generation in both languages.

\begin{table}[t]
\vspace*{6pt}
\centering
\small
\setlength{\tabcolsep}{5pt}
\begin{tabular}{lccc}
\toprule
Model & WER(\%)$\downarrow$ & SIM$\uparrow$ & UTMOS$\uparrow$ \\
\midrule
\multicolumn{4}{c}{Seed-TTS \textit{test-en}} \\
\midrule
Llasa-1B-80k & 3.71 & 0.54 & 4.06 \\
Llasa-1B-160k & 3.60 & 0.56 & 4.05 \\
Llasa-1B-250k & 2.99 & 0.57 & 4.07 \\
Spark-TTS & 1.98 & 0.58 & 3.94 \\
\midrule
Ours & \textbf{1.87} & \textbf{0.59} & \textbf{4.11} \\
\midrule
\multicolumn{4}{c}{Seed-TTS \textit{test-zh}} \\
\midrule
Llasa-1B-80k & 2.69 & 0.65 & 3.27 \\
Llasa-1B-160k & 2.22 & 0.66 & 3.28 \\
Llasa-1B-250k & 1.89 & \textbf{0.67} & 3.28 \\
Spark-TTS & \textbf{1.20} & \textbf{0.67} & 3.27 \\
\midrule
Ours & 1.72 & 0.65 & \textbf{3.30} \\
\bottomrule
\end{tabular}
\caption{LLM-based speech generation performance on the Seed-TTS-Eval benchmark.}
\label{tab:tts-generation}
\end{table}

\begin{table*}[!t]
\centering
\scriptsize
\setlength{\tabcolsep}{6pt}
\resizebox{\textwidth}{!}{
\begin{tabular}{lccccccc}
\toprule
\multirow{2}{*}{Model} &
\multirow{2}{*}{Params} &
\multicolumn{2}{c}{Encode} &
\multicolumn{2}{c}{Decode} &
\multicolumn{2}{c}{Total} \\
\cmidrule(lr){3-4}\cmidrule(lr){5-6}\cmidrule(lr){7-8}
& &
MACs$\downarrow$ & RTF$\downarrow$ &
MACs$\downarrow$ & RTF$\downarrow$ &
MACs$\downarrow$ & RTF$\downarrow$ \\
\midrule
EnCodec & 15M & \underline{2.80G} & 0.005 & \textbf{2.80G} & 0.004 & \textbf{5.60G} & 0.009 \\
DAC & 74M & 12.27G & 0.005 & 43.22G & 0.003 & 55.49G & 0.008 \\
BigCodec & 159M & 25.52G & 0.008 & 35.56G & 0.006 & 61.08G & 0.014 \\
Mimi & 79M & \textbf{2.72G} & \textbf{0.003} & \underline{8.60G} & \textbf{0.001} & \underline{11.32G} & \textbf{0.004} \\
DualCodec & 664M & 31.83G & 0.015 & 9.15G & 0.004 & 40.98G & 0.019 \\
XY-Tokenizer & 520M & 63.59G & 0.005 & 18.43G & \underline{0.002} & 82.02G & \underline{0.007} \\
BiMTokenizer (Ours) & 253M & 4.79G & \underline{0.004} & 10.03G & 0.003 & 14.82G & \underline{0.007} \\
\bottomrule
\end{tabular}}
\caption{Codec-level computational efficiency on LibriSpeech \textit{test-clean}. MACs are computed for a 1-second audio input, and RTF denotes real-time factor. Total costs sum encoding and decoding. Lower values indicate higher efficiency.}
\label{tab:efficiency}
\end{table*}

\afterpage{\afterpage{%
\begin{figure}[!t]
  \centering
  \includegraphics[width=0.86\columnwidth]{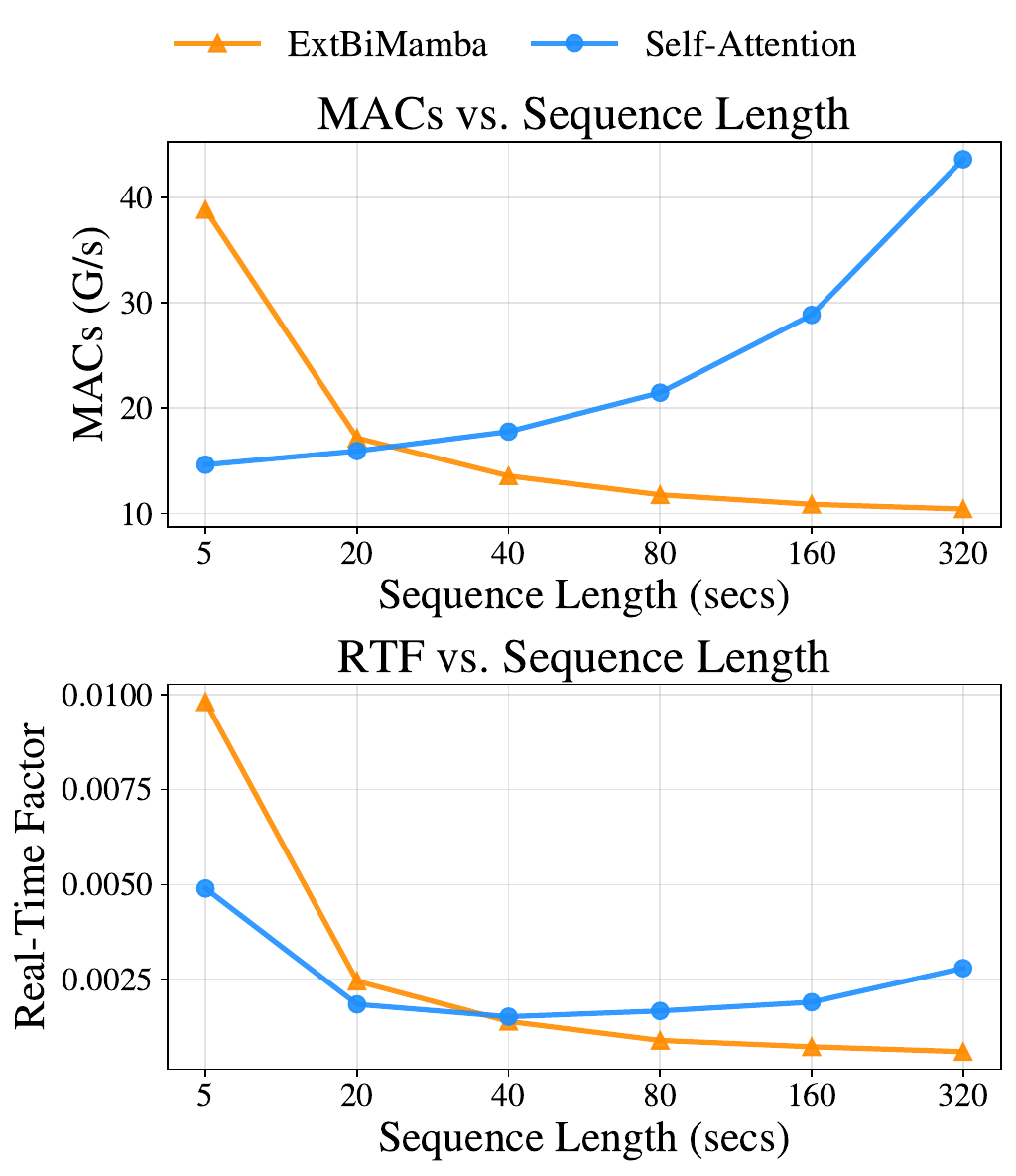}
  \caption{Length-scaling comparison between the BiMamba backbone and the bidirectional Transformer counterpart with rotary position embeddings (RoPE).}
  \label{fig:mamba-vs-transformer-efficiency}
\end{figure}
}}

\subsection{Further Analysis}
\paragraph{Computational Efficiency.}

We assess computational efficiency at both the codec and backbone levels. At the codec level, Table~\ref{tab:efficiency} shows that BiMTokenizer contains 253M parameters and requires 14.82G MACs, compared with 664M parameters and 40.98G MACs for DualCodec and 520M parameters and 82.02G MACs for XY-Tokenizer. Relative to these two dual-tower codecs, BiMTokenizer reduces total MACs by 63.8\% and 81.9\%, respectively, with particularly low encoding cost for large-scale corpus tokenization. The total RTF of BiMTokenizer is 0.007, matching that of XY-Tokenizer and remaining substantially below the 0.019 RTF of DualCodec. At the backbone level, Figure~\ref{fig:mamba-vs-transformer-efficiency} shows that BiMamba gains efficiency as the input duration grows, whereas bidirectional RoPE-based self-attention~\citep{su2024roformer} incurs steeper growth in computation and runtime. Together, these results show that the single-tower design reduces end-to-end computational overhead and scales favorably to long-form audio. Detailed measurement settings and additional analysis are provided in Appendix~\ref{sec:appendix-efficiency}.

\paragraph{Component Analysis.}
Table~\ref{tab:ablation} in Appendix~\ref{sec:appendix-ablation} isolates the backbone and quantizer effects without semantic supervision. Replacing BiMamba with bidirectional self-attention raises WER from 2.55\% to 4.15\% and lowers PESQ from 3.56/2.99 to 3.28/2.77. Replacing RSLQ with GroupFSQ or RVQ likewise degrades all reconstruction metrics. Thus, BiMamba benefits both efficiency and reconstruction, while RSLQ outperforms the alternative quantizers at the same bitrate.

Teacher-supervised ablations reveal a semantic--acoustic trade-off. For SenseVoice at $\lambda_\text{sem}{=}15$, reconstruction alignment lowers WER from 2.58\% to 2.53\% and raises AudioMNIST from 97.79 to 98.02, whereas increasing the semantic weight to 20 improves SLURP at the cost of SIM and PESQ. Whisper shows the same trend, with $\lambda_\text{sem}{=}15$ giving the best overall acoustic performance. We set $\lambda_\text{sem}{=}15$ and $\lambda_\text{align}{=}1$ in both final variants.

\section{Conclusion}
\label{sec:conclusion}

We presented BiMTokenizer, a 1.1\,kbps single-tower speech codec built on a bidirectional state-space backbone and Residual Spherical Leech Quantization. Experiments on reconstruction and semantic benchmarks show that BiMTokenizer matches or surpasses recent dual-tower codecs while using far fewer parameters. These results offer a counterpoint to the recent shift toward dual-tower designs, suggesting that careful redesign of a single-tower codec's shared backbone and bottleneck remains an equally viable path for low-bitrate speech tokenization.

\FloatBarrier

\section*{Acknowledgments}
This work was partially supported by the Guangxi Science and Technology Major Program under Grant No.~AA24206067.
A large language model was used solely to assist with language polishing and improving the clarity of writing.

\section*{Limitations}
Several limitations remain. First, RSLQ uses a fixed 196,560-entry codebook. The resulting vocabulary provides high capacity and separation but enlarges the prediction space for autoregressive TTS, potentially requiring more training data and model capacity to learn infrequent codec symbols. This capacity--predictability trade-off warrants further study. Second, the current codec combines convolutional encoder/decoder modules with Mamba blocks, resulting in a heterogeneous CNN--Mamba architecture. An architecturally homogeneous, fully Mamba-based codec is a promising direction for future work. Finally, the bidirectional backbone requires future frames, limiting the codec to offline tokenization. Causal or chunk-wise variants are needed for low-latency streaming.

\vspace*{\fill}
\section*{Ethical Considerations}
This work studies a speech codec at the representation level and is intended for research purposes. Ethical considerations such as privacy, bias, and misuse are primarily determined by downstream
applications rather than the codec itself. We encourage responsible use according to ethical guidelines.

\clearpage
\bibliography{custom}

\clearpage

\appendix
\setlength{\parskip}{0pt plus 4pt}

\section{Details of Architecture}
\label{sec:appendix-method}

\subsection{Encoder}

The input waveform is resampled to 16 kHz and converted to an 80-channel mel-spectrogram using a 25 ms window and a 10 ms hop. The trainable codec encoder follows a Whisper-style convolutional front-end~\citep{radford2023whisper}: two 1D convolutional layers with kernel size 3 project the input to 768 dimensions and reduce the frame rate from 100 Hz to 50 Hz. The resulting sequence is processed by 8 bidirectional Mamba-1 blocks~\citep{gu2023mamba,zhang2025mambaspeech} with hidden dimension 768, state dimension 16, convolution width 4, expansion factor 2, RMSNorm, and a 1536-dimensional SwiGLU feed-forward layer.

\subsection{Down/Upsampling}

The quantizer operates at 12.5 Hz. The downsampler stacks every four adjacent 50 Hz encoder frames along the channel dimension and maps the resulting representation to a 128-dimensional latent sequence $z_e$ through weight-normalized 1D projections and residual convolutional units with dilations 1, 3, and 9, following common neural audio codec designs~\citep{zeghidour2021soundstream,defossez2022encodec}.

The upsampler mirrors this module. It projects the quantized 128-dimensional latent back to $768\times4$ channels, applies the same residual convolutional units, and unstacks frames to recover a 50 Hz sequence before decoding. No attention-based projection is used in either sampling module.

\subsection{Quantizer}

The bottleneck uses 5 RSLQ levels with a fixed spherical Leech codebook of 196,560 entries per level~\citep{zhao2026spherical,ConwayS88}. The first level is used as the semantic SLQ level, and the remaining four levels form the residual acoustic path. The main text gives the RSLQ computation in Sec.~\ref{sec:rslq}.

\subsection{Decoder}

The decoder mirrors the encoder. It applies 8 bidirectional Mamba-1 blocks to the upsampled 50 Hz sequence and reconstructs an 80-channel mel-spectrogram with two transposed 1D convolutional layers. A jointly trained Vocos vocoder~\citep{siuzdak2024vocos} with 12 layers, hidden dimension 512, intermediate dimension 4096, FFT size 640, and hop size 160 converts the reconstructed mel-spectrogram into a 16 kHz waveform.

\subsection{Semantic Teacher}

The semantic branch is only used during training. We instantiate the frozen teacher as either Whisper-small~\citep{radford2023whisper} or SenseVoice-small~\citep{funaudiollm2024} and use its representations for semantic distillation and reconstruction alignment, as described in Sec.~\ref{sec:semantic-supervision}. The teacher encoder and the auxiliary projection/alignment paths are removed at inference, leaving the acoustic codec as a single-tower encoder--RSLQ--decoder system.

\section{RSLQ Codebook Behavior}
\label{app:rslq}

\begin{figure*}[t]
  \centering
  \includegraphics[width=0.86\textwidth]{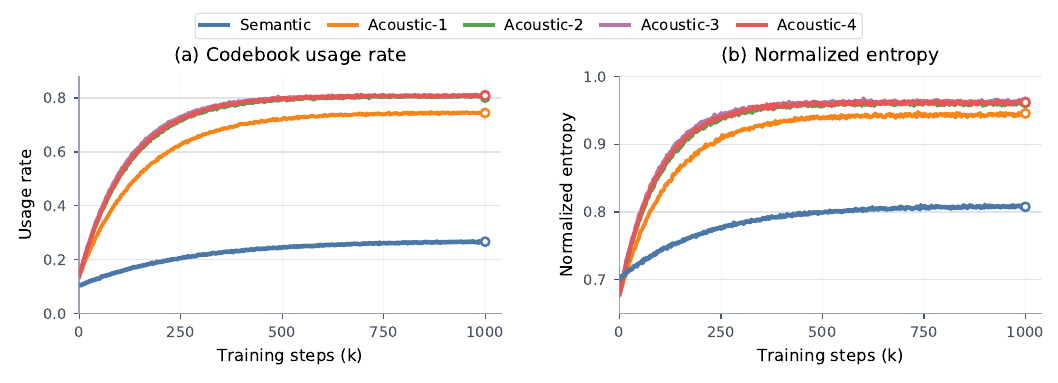}
  \caption{Validation codebook behavior of the five RSLQ levels. The semantic level shows lower but stable usage and normalized entropy under semantic supervision, while the four residual acoustic levels maintain higher utilization and entropy throughout training.}
  \label{fig:rslq-codebook-behavior}
\end{figure*}

We analyze the validation-time codebook behavior of the five RSLQ levels to verify that the fixed Leech codebook~\citep{zhao2026spherical,ConwayS88} is effectively used under the proposed split-residual design. Figure~\ref{fig:rslq-codebook-behavior} reports the codebook usage rate and normalized entropy during training. The first level, which receives semantic supervision, maintains a lower but stable usage rate and a normalized entropy of about 0.81. This pattern is expected because semantic distillation encourages a more concentrated token distribution over linguistically relevant regions of the codebook, rather than uniformly spreading probability mass across all entries. In contrast, the four residual acoustic levels reach substantially higher usage rates and normalized entropy values above 0.94, indicating broad utilization of the fixed lattice codebook for speaker, prosodic, and fine acoustic residual information.

These results support the intended division of labor in RSLQ. The semantic level forms a compact supervised allocation, while the acoustic residual levels preserve high-entropy code usage. Importantly, all levels remain stable after convergence, suggesting that the fixed Leech lattice avoids the severe under-utilization and collapse commonly associated with learnable VQ/RVQ codebooks~\citep{vandenoord2017vqvae,zeghidour2021soundstream,defossez2022encodec,kumar2023dac}.

\section{Training Objective Details}
\label{sec:appendix-objective}

To accelerate training, we pre-extract semantic teacher representations for the full LibriSpeech training corpus~\citep{panayotov2015librispeech} and store them offline. For the Whisper-small variant~\citep{radford2023whisper}, we store the 50 Hz encoder features directly. For the SenseVoice-small variant~\citep{funaudiollm2024}, the teacher produces 16.67 Hz features; we linearly interpolate them to the 12.5 Hz codec token rate before storage, so that semantic distillation can be applied at the RSLQ bottleneck resolution. During generator training, both teacher-supervised variants use the same loss coefficients: $\lambda_{\mathrm{sem}}=15$, $\lambda_{\mathrm{align}}=1$, $\lambda_{\mathrm{rec}}=15$, $\lambda_{\mathrm{adv}}=1$, and $\lambda_{\mathrm{feat}}=1$. The frozen teacher features are used only for auxiliary semantic supervision and reconstruction alignment, and no teacher branch is used at inference.

\setcounter{dbltopnumber}{2}

\begin{table*}[t]
  \centering
  \scriptsize
  \setlength{\tabcolsep}{2.4pt}
  \resizebox{\textwidth}{!}{
  \begin{tabular}{lcccccccccccc}
    \toprule
    \multirow{2}{*}{Model} & Codebook & \multirow{2}{*}{bps} &
    Frame & \multirow{2}{*}{$N_q$} &
    Single & \multirow{2}{*}{Param} &
    \multirow{2}{*}{SIM$\uparrow$} & \multirow{2}{*}{STOI$\uparrow$} &
    PESQ & PESQ &
    \multirow{2}{*}{UTMOS$\uparrow$} &
    \multirow{2}{*}{WER{(\%)}$\downarrow$} \\
    & Size & & Rate & & Tower & & & &
    NB$\uparrow$ & WB$\uparrow$ & & \\
    \midrule
    Ground Truth & -- & -- & -- & -- & -- & -- & 1.00 & 1.00 & 4.55 & 4.64 & 3.50 & 4.59 \\
    \midrule
    EnCodec (24 kHz) & 1024 & 1500 & 75 & 2 & \checkmark & 15M & 0.59 & 0.83 & 1.96 & 1.58 & 1.48 & 17.19 \\
    DAC (16 kHz) & 1024 & 1500 & 50 & 3 & \checkmark & 74M & 0.45 & 0.79 & 1.61 & 1.26 & 1.42 & 22.31 \\
    SpeechTokenizer (16 kHz) & 1024 & 1000 & 50 & 2 & \checkmark & 103.7M & 0.32 & 0.74 & 1.51 & 1.23 & 1.93 & 14.71 \\
    BigCodec (16 kHz) & 8192 & 1040 & 80 & 1 & \checkmark & 159M & 0.82 & 0.91 & 3.03 & 2.44 & 3.55 & 8.29 \\
    Mimi (24 kHz) & 2048 & 1100 & 12.5 & 8 & \checkmark & 79M & 0.71 & 0.88 & 2.64 & 2.12 & 3.07 & 9.41 \\
    XCodec (16 kHz) & 1024 & 1000 & 50 & 2 & $\times$ & 160M & 0.67 & 0.85 & 2.50 & 1.99 & 3.66 & 7.02 \\
    XCodec2.0 (16 kHz) & 65536 & 800 & 50 & 1 & $\times$ & 820M & 0.81 & 0.90 & 2.83 & 2.26 & 3.64 & 6.85 \\
    DualCodec (24 kHz) & 16384/4096 & 1075 & 12.5 & 1/6 & $\times$ & 664M & 0.83 & 0.91 & 3.07 & 2.54 & 3.65 & 6.26 \\
    XY-Tokenizer (16 kHz) & 1024 & 1000 & 12.5 & 8 & $\times$ & 520M & \underline{0.84} & 0.90 & 2.88 & 2.28 & 3.46 & \underline{6.19} \\
    SimWhisper-Codec (16 kHz) & 2016 & 1100 & 12.5 & 8 & \checkmark & 291M & 0.81 & \underline{0.92} & 3.12 & 2.56 & 3.49 & 7.61 \\
    \midrule
    BiMTokenizer-Whisper (Ours, 16 kHz) & 196560 & 1100 & 12.5 & 5 & \checkmark & 253M & \textbf{0.85} & \textbf{0.93} & \textbf{3.38} & \textbf{2.83} & \underline{3.75} & \textbf{6.10} \\
    BiMTokenizer-SenseVoice (Ours, 16 kHz) & 196560 & 1100 & 12.5 & 5 & \checkmark & 253M & 0.83 & \underline{0.92} & \underline{3.25} & \underline{2.63} & \textbf{3.76} & 6.32 \\
    \bottomrule
  \end{tabular}}
  \caption{Speech reconstruction comparison on the noisier LibriSpeech \textit{test-other} set. Best codec results are in bold, and second-best results are underlined.}
  \label{tab:test-other}
\end{table*}

\begin{table*}[t]
  \centering
  \scriptsize
  \setlength{\tabcolsep}{2.6pt}
  \resizebox{\textwidth}{!}{
  \begin{tabular}{lccccccccccc}
    \toprule
    \multirow{2}{*}{Model} & \multirow{2}{*}{kbps} &
    \multicolumn{5}{c}{SeedTTS-ZH} &
    \multicolumn{5}{c}{SeedTTS-EN} \\
    \cmidrule(lr){3-7}\cmidrule(lr){8-12}
    & &
    PESQ-NB$\uparrow$ & PESQ-WB$\uparrow$ & SIM$\uparrow$ & STOI$\uparrow$ & UTMOS$\uparrow$ &
    PESQ-NB$\uparrow$ & PESQ-WB$\uparrow$ & SIM$\uparrow$ & STOI$\uparrow$ & UTMOS$\uparrow$ \\
    \midrule
    Ground Truth & -- & 4.55 & 4.64 & 1.00 & 1.00 & 2.78 & 4.55 & 4.64 & 1.00 & 1.00 & 3.53 \\
    \midrule
    BigCodec (16 kHz) & 1.04 & 2.88 & 2.26 & 0.80 & 0.91 & 2.90 & 2.80 & 2.21 & 0.80 & 0.91 & 3.59 \\
    XCodec2.0 (16 kHz) & 0.80 & 2.85 & 2.23 & 0.85 & 0.90 & 2.89 & 2.67 & 2.09 & 0.81 & 0.90 & 3.69 \\
    XY-Tokenizer (16 kHz) & 1.00 & 2.98 & 2.31 & \textbf{0.88} & 0.91 & 2.78 & 2.75 & 2.18 & \textbf{0.84} & 0.90 & 3.59 \\
    SimWhisper-Codec (16 kHz) & 1.10 & 3.08 & 2.44 & 0.82 & 0.92 & 2.99 & 2.89 & 2.28 & 0.80 & 0.92 & 3.61 \\
    \midrule
    BiMTokenizer-Whisper (Ours, 16 kHz) & 1.10 & \textbf{3.30} & \textbf{2.66} & 0.85 & \textbf{0.93} & 3.20 & \textbf{3.13} & \textbf{2.51} & \textbf{0.84} & \textbf{0.93} & 3.89 \\
    BiMTokenizer-SenseVoice (Ours, 16 kHz) & 1.10 & 3.15 & 2.49 & 0.82 & 0.92 & \textbf{3.23} & 2.93 & 2.28 & 0.80 & 0.92 & \textbf{3.91} \\
    \bottomrule
  \end{tabular}}
  \caption{Speech reconstruction comparison on Seed-TTS-Eval dataset. ZH/EN split results are reported in the same row for each model. Best codec results are in bold.}
  \label{tab:seedtts}
\end{table*}

\subsection{Semantic Projector}
\label{sec:appendix-semantic-projector}

The lightweight projector $P_{\psi}$ depends on the frozen semantic teacher. For the Whisper-small variant~\citep{radford2023whisper}, the teacher encoder produces 50 Hz features, while the semantic RSLQ level operates at 12.5 Hz. We therefore implement $P_{\psi}$ with the same linear upsampling module used in the codec, converting the 12.5 Hz semantic sequence to 50 Hz before aligning it with Whisper encoder features. For the SenseVoice-small variant~\citep{funaudiollm2024}, $P_{\psi}$ is a single linear layer that maps the semantic RSLQ embedding to the SenseVoice feature dimension.

\subsection{Reconstruction Loss}

We use a multi-scale mel-spectrogram reconstruction loss, following common practice in neural speech generation and codec training~\citep{kong2020hifigan,defossez2022encodec}. For each STFT scale $k\in\{5,\ldots,11\}$, let $M_k(\cdot)$ denote the mel-spectrogram computed with FFT size $2^k$. Given the original audio $x$ and reconstructed audio $\hat{x}$, the reconstruction loss is
\begin{equation}
  \mathcal{L}_{\mathrm{rec}}=
  \sum_{k=5}^{11}\|M_k(x)-M_k(\hat{x})\|_1.
\end{equation}
This objective focuses reconstruction learning on perceptually relevant spectral structure. We do not use an additional time-domain waveform L1 term.

\subsection{Adversarial and Feature-Matching Losses}

We use a multi-period discriminator (MPD)~\citep{kong2020hifigan} and a multi-scale short-time Fourier transform discriminator (MS-STFTD)~\citep{defossez2022encodec}. Let $D_i(\cdot)$ denote the output of the $i$-th discriminator and let $N$ be the number of discriminators. The discriminator is optimized with the least-squares GAN objective~\citep{mao2017lsgan}:
\begin{equation}
  \mathcal{L}_{D}=
  \frac{1}{N}\sum_{i=1}^{N}
  \left[(D_i(x)-1)^2 + D_i(\hat{x})^2\right].
\end{equation}
The generator adversarial loss is
\begin{equation}
  \mathcal{L}_{\mathrm{adv}}=
  \frac{1}{N}\sum_{i=1}^{N}(D_i(\hat{x})-1)^2.
\end{equation}
We also apply feature matching on intermediate discriminator activations. Let $D_i^{j}(\cdot)$ denote the $j$-th layer feature map from discriminator $D_i$, let $K$ be the number of feature layers, and let $\epsilon$ be a small constant for numerical stability. The feature-matching loss is
\begin{equation}
  \mathcal{L}_{\mathrm{feat}}=
  \frac{1}{N K}
  \sum_{i=1}^{N}\sum_{j=1}^{K}
  \frac{\|D_i^{j}(x)-D_i^{j}(\hat{x})\|_1}
  {\|D_i^{j}(x)\|_1+\epsilon},
\end{equation}
which stabilizes adversarial training by matching real and reconstructed audio in discriminator feature space. Since RSLQ has no learnable codebook, no VQ codebook loss, commitment loss, or entropy regularization is added.

\section{Evaluation Details}
\label{sec:appendix-evaluation}

For model-level comparison on LibriSpeech \textit{test-clean}~\citep{panayotov2015librispeech}, we report codebook size, bitrate in bits per second (bps), frame rate, number of quantizers ($N_q$), single-tower architecture status, and parameter count. Speech intelligibility is measured by Short-Time Objective Intelligibility (STOI)~\citep{taal2010stoi} and word error rate (WER). WER transcriptions are obtained with a HuBERT-based ASR model~\citep{hsu2021hubert}.\footnote{\url{https://huggingface.co/facebook/hubert-large-ls960-ft}}

Acoustic quality is measured by narrow-band and wide-band Perceptual Evaluation of Speech Quality (PESQ-NB/PESQ-WB)~\citep{rix2001pesq}, UTMOS~\citep{saeki2022utmos}, and the Virtual Speech Quality Objective Listener (ViSQOL)~\citep{chinen2020visqol}. We also report speaker similarity (SIM), computed as the cosine similarity between speaker embeddings extracted from the original and reconstructed speech using a pretrained speaker verification model.\footnote{\url{https://github.com/microsoft/UniSpeech/tree/main/downstreams/speaker_verification}}

\section{Details of LLM-Based Speech Generation}
\label{sec:appendix-tts}

\subsection{Speech Generation Model and Training Details}

For autoregressive speech generation, we retrain a TTS-oriented variant of BiMTokenizer on the English and Chinese subsets of Emilia~\citep{he2025emilia}, totaling approximately 96.7K hours of speech (46.8K English and 49.9K Chinese). To make token-level language-model training tractable, we subsample the original spherical Leech codebook to 2,048 entries at every quantization level. The tokenizer uses eight quantization levels: one semantic level followed by seven acoustic residual levels. At a frame rate of 12.5 Hz, each level contributes 11 bits per frame, yielding a total bitrate of $12.5\times8\times11=1{,}100$ bps. We adopt Qwen3-0.6B~\citep{yang2025qwen3technicalreport} as the language-model backbone and follow the autoregressive training formulation of Qwen3-TTS~\citep{hu2026qwen3ttstechnicalreport}.

During training, text tokens and speech tokens are concatenated along the temporal dimension, and the autoregressive loss is computed only on the speech tokens. The model is trained on the VoxBox dataset proposed in Spark-TTS~\citep{wang2025spark}. We use a maximum token budget of 131,072 per training step, train the model for 400k steps, and set the maximum learning rate to $2\times10^{-4}$.

\subsection{Evaluation Protocol}

We evaluate zero-shot speech generation on the Seed-TTS-Eval benchmark~\citep{anastassiou2024seedtts}, including both the English (\textit{test-en}) and Chinese (\textit{test-zh}) subsets. We report word error rate (WER), speaker similarity (SIM), and UTMOS. During inference, the model is conditioned on the prompt text, reference speech, and target text, and autoregressively generates the target speech tokens, which are then decoded into waveforms by the BiMTokenizer decoder.

\section{Reconstruction under Noisy and Out-of-Distribution Conditions}
\label{sec:appendix-robustness}

\paragraph{Noisy Conditions.} To further evaluate robustness under more challenging acoustic conditions, we conduct additional reconstruction experiments on the full LibriSpeech \textit{test-other} set~\citep{panayotov2015librispeech}. Compared with \textit{test-clean}, this subset contains noisier and more variable speech, making it a useful setting for testing whether a codec can preserve intelligibility, speaker information, and perceptual quality beyond clean read speech.

The ground-truth recordings already show a clear degradation from \textit{test-clean}: UTMOS~\citep{saeki2022utmos} decreases from 4.09 to 3.50, and WER increases from 2.16\% to 4.59\%. Most codecs therefore exhibit lower STOI~\citep{taal2010stoi} and PESQ~\citep{rix2001pesq} scores and higher WER under this setting. Nevertheless, BiMTokenizer remains strong on noisy speech. The Whisper-supervised variant achieves the best codec result on SIM, STOI, PESQ-NB, PESQ-WB, and WER, showing stronger reconstruction quality and intelligibility than the competing codecs. The SenseVoice-supervised variant obtains the highest UTMOS, although its gains over the Whisper-supervised variant are limited to this metric.

\paragraph{Out-of-Distribution} To further assess generalization beyond the training distribution, we evaluate BiMTokenizer on Seed-TTS-Eval~\citep{anastassiou2024seedtts}, which covers both English (\textit{seedtts-test-en}) and Mandarin Chinese (\textit{seedtts-test-zh}). Since our codec is trained exclusively on English LibriSpeech~\citep{panayotov2015librispeech}, seedtts-test-en probes generalization across speakers, recording conditions, and content style, whereas seedtts-test-zh additionally probes \textit{cross-lingual} generalization to a language entirely unseen during training. Following common practice in speech codec evaluation on this benchmark, we report PESQ-NB/PESQ-WB~\citep{rix2001pesq}, SIM, STOI~\citep{taal2010stoi}, and UTMOS~\citep{saeki2022utmos}.

As shown in Table~\ref{tab:seedtts}, BiMTokenizer attains the strongest results on the majority of metrics on both subsets. BiMTokenizer-Whisper leads on PESQ-NB, PESQ-WB, and STOI on both seedtts-test-en (3.13, 2.51, 0.93) and seedtts-test-zh (3.30, 2.66, 0.93), and ties XY-Tokenizer~\citep{gong2025xytokenizer} for the best SIM on seedtts-test-en (0.84). BiMTokenizer-SenseVoice obtains the highest UTMOS on both subsets (3.91 on en, 3.23 on zh). The only metric on which BiMTokenizer does not lead is SIM on seedtts-test-zh, where XY-Tokenizer reaches 0.88 versus our 0.85; this is consistent with XY-Tokenizer being trained on a corpus that includes Mandarin speakers, whereas our model has had no exposure to Mandarin speaker characteristics during training, and the gap is small in absolute terms. Importantly, the codec-internal ranking is preserved across the two languages, indicating that the learned representations encode acoustic and phonetic regularities that transfer beyond the training language rather than fitting language-specific patterns of LibriSpeech English.

These results indicate that the proposed single-tower architecture is not only effective on clean speech but also robust under more difficult acoustic conditions and generalizable to out-of-distribution and even cross-lingual data. Compared with recent dual-tower codecs~\citep{li2025dualcodec,gong2025xytokenizer,chen2025sac}, BiMTokenizer achieves stronger results on most reconstruction metrics while retaining a single-tower architecture. This further supports our main claim that the single-tower route has not been exhausted, and that improving the temporal backbone and quantization bottleneck can substantially strengthen low-bitrate speech coding.

\begin{table*}[t]
  \centering
  \scriptsize
  \setlength{\tabcolsep}{4pt}
  \renewcommand{\arraystretch}{1.08}
  \resizebox{\textwidth}{!}{
  \begin{tabular}{lcccllcccccc}
    \toprule
    \multirow{2}{*}{Group} & \multirow{2}{*}{Teacher} & \multirow{2}{*}{$\lambda_{\mathrm{sem}}$} & \multirow{2}{*}{$\lambda_{\mathrm{align}}$} & \multirow{2}{*}{Backbone} & \multirow{2}{*}{Quantizer} & \multicolumn{4}{c}{Reconstruction} & \multicolumn{2}{c}{Understanding} \\
    \cmidrule(lr){7-10}\cmidrule(lr){11-12}
    & & & & & & SIM$\uparrow$ & UTMOS$\uparrow$ & PESQ$\uparrow$ & WER$\downarrow$ & SLURP$\uparrow$ & AM$\uparrow$ \\
    \midrule
    \multirow{4}{*}{Core Design}
    & -- & 0 & 0 & BiMamba & RSLQ & \textbf{0.88} & \textbf{4.12} & \textbf{3.56}/\textbf{2.99} & \textbf{2.55} & \textbf{7.94} & \textbf{78.26} \\
    & -- & 0 & 0 & Bi-Self-Attn & RSLQ & \underline{0.85} & 3.90 & 3.28/2.77 & 4.15 & 7.62 & 75.18 \\
    & -- & 0 & 0 & BiMamba & GroupFSQ & 0.84 & \underline{4.03} & \underline{3.42}/\underline{2.84} & \underline{2.69} & 7.75 & 75.31 \\
    & -- & 0 & 0 & BiMamba & RVQ & 0.84 & 4.00 & \underline{3.42}/2.75 & 2.75 & \underline{7.81} & \underline{76.26} \\
    \midrule
    \multirow{4}{*}{SenseVoice}
    & SenseVoice & 0 & 1 & BiMamba & RSLQ & \textbf{0.86} & \underline{4.15} & \textbf{3.50}/\textbf{2.95} & 2.56 & 8.58 & 81.25 \\
    & SenseVoice & 15 & 0 & BiMamba & RSLQ & 0.84 & \textbf{4.18} & \underline{3.46}/\underline{2.89} & 2.58 & 18.22 & \underline{97.79} \\
    & SenseVoice & 15 & 1 & BiMamba & RSLQ & \underline{0.85} & \textbf{4.18} & 3.45/2.85 & \textbf{2.53} & \underline{18.48} & \textbf{98.02} \\
    & SenseVoice & 20 & 1 & BiMamba & RSLQ & 0.82 & \underline{4.15} & 3.42/2.78 & \underline{2.55} & \textbf{19.90} & 97.46 \\
    \midrule
    \multirow{4}{*}{Whisper}
    & Whisper & 5 & 1 & BiMamba & RSLQ & \underline{0.86} & 4.07 & 3.46/2.86 & \underline{2.51} & 10.89 & 84.65 \\
    & Whisper & 10 & 1 & BiMamba & RSLQ & \underline{0.86} & 4.10 & 3.48/2.92 & \underline{2.51} & 12.24 & \underline{86.28} \\
    & Whisper & 15 & 1 & BiMamba & RSLQ & \textbf{0.87} & \textbf{4.21} & \textbf{3.56}/\textbf{3.03} & \textbf{2.44} & \underline{12.51} & 86.20 \\
    & Whisper & 20 & 1 & BiMamba & RSLQ & \textbf{0.87} & \underline{4.18} & \underline{3.52}/\underline{2.99} & 2.52 & \textbf{13.68} & \textbf{87.11} \\
    \bottomrule
  \end{tabular}}
  \caption{Component analysis on reconstruction and semantic representation. The core-design block removes semantic supervision and first reports the full BiMamba+RSLQ reconstruction-only setting, followed by backbone and quantizer replacements. The SenseVoice block studies semantic distillation, reconstruction alignment, and distillation weights under the SenseVoice teacher. The Whisper block varies the distillation weight with reconstruction alignment enabled. PESQ reports PESQ-NB/PESQ-WB. Best and second-best values are marked within each block for each metric. AM denotes AudioMNIST.}
  \label{tab:ablation}
\end{table*}

\section{Computational Efficiency Details}
\label{sec:appendix-efficiency}

We provide two complementary efficiency analyses. The first isolates the sequence backbone by comparing BiMamba~\citep{gu2023mamba,zhang2025mambaspeech} with a bidirectional Transformer counterpart~\citep{vaswani2017attention}, and the second measures the complete encoding and decoding cost of each codec. We report Multiply-Accumulate operations (MACs) for arithmetic complexity and Real-Time Factor (RTF) for wall-clock inference speed relative to audio duration. Following common neural audio codec evaluation practice~\citep{zeghidour2021soundstream,defossez2022encodec,kumar2023dac}, MACs are computed for a 1-second audio input and reported in G/s. We use PyFlops for supported modules and manually calculate modules not supported by PyFlops, such as state-space model (SSM) blocks~\citep{gu2023mamba}. All measurements are conducted with batch size 1 on a single NVIDIA H100 GPU. For the codec-level comparison in Table~\ref{tab:efficiency}, both MACs and RTF are measured on LibriSpeech test-clean~\citep{panayotov2015librispeech}.

Figure~\ref{fig:mamba-vs-transformer-efficiency} isolates the computational effect of the sequence backbone. The BiMamba~\citep{zhang2025mambaspeech} and Transformer~\citep{vaswani2017attention} variants share the same codec encoder front-end, linear downsampling and upsampling modules, RSLQ bottleneck, decoder, and vocoder. The only architectural change is that each bidirectional Mamba block is replaced by a bidirectional self-attention block with RoPE~\citep{su2024roformer}. This controlled setting ensures that the comparison reflects the backbone design rather than differences in temporal context, quantization, or decoder structure. MACs and RTF are measured for audio inputs of 5, 20, 40, 80, 160, and 320 seconds, with all non-backbone codec components kept unchanged.

Table~\ref{tab:efficiency} reports codec-level efficiency. Parameter count is listed once per model, while encoding and decoding costs are reported separately. Encoding corresponds to waveform-to-token conversion, and decoding corresponds to waveform reconstruction from the discrete tokens. Total MACs and total RTF are obtained by summing the corresponding encoding and decoding values, which reflects the full inference cost of tokenization followed by reconstruction.

MACs and RTF provide complementary measures. MACs quantify arithmetic complexity, whereas RTF also depends on hardware and operator implementation. Reporting both therefore separates the model's computational footprint from its realized wall-clock speed. Section~\ref{sec:results} discusses the quantitative comparisons and their implications.

\section{Component Analysis Details}
\label{sec:appendix-ablation}

Table~\ref{tab:ablation} is organized to isolate component choices from supervision choices. The core-design block removes semantic supervision and evaluates every variant at the same 1.1\,kbps bitrate. In the backbone ablation, only the BiMamba blocks are replaced by bidirectional RoPE-based self-attention~\citep{vaswani2017attention,su2024roformer}; all other codec components remain unchanged. In the quantizer ablations, GroupFSQ~\citep{casanova2025lowframe} uses 8 groups with per-group levels $[8,7,6,6]$, while RVQ uses 8 quantizer layers with a codebook size of 2048 per layer, following standard neural codec practice~\citep{zeghidour2021soundstream,defossez2022encodec,kumar2023dac}. These controlled settings ensure that the observed differences can be attributed to the backbone or quantizer rather than to bitrate or semantic supervision.

Within the reconstruction-only block, BiMamba with RSLQ achieves the strongest result on every reconstruction metric. Replacing BiMamba with self-attention primarily affects intelligibility and spectral fidelity, while replacing RSLQ with GroupFSQ or RVQ produces smaller but consistent degradations across SIM, UTMOS, PESQ, and WER. The downstream understanding scores in this block are reported for completeness, but they should not be interpreted as measures of semantic supervision because none of these variants uses a semantic teacher.

The teacher-specific blocks retain the BiMamba+RSLQ architecture and vary only the semantic objectives. In the SenseVoice block~\citep{funaudiollm2024}, introducing semantic distillation at $\lambda_{\mathrm{sem}}=15$ increases SLURP accuracy from 8.58 to 18.48 and AudioMNIST accuracy from 81.25 to 98.02, while preserving competitive reconstruction quality. Comparing the two settings with $\lambda_{\mathrm{sem}}=15$ further isolates the contribution of reconstruction alignment. Enabling this objective improves WER from 2.58\% to 2.53\% and AudioMNIST accuracy from 97.79 to 98.02. Increasing $\lambda_{\mathrm{sem}}$ to 20 yields the highest SLURP accuracy, but reduces SIM and PESQ, indicating that excessive semantic supervision compromises acoustic preservation.

The Whisper block~\citep{radford2023whisper} exhibits a similar trend. Increasing $\lambda_{\mathrm{sem}}$ from 5 to 15 improves reconstruction, intelligibility, and SLURP performance. A further increase to 20 produces the best SLURP and AudioMNIST scores in this block, but slightly degrades UTMOS, PESQ, and WER. We therefore use $\lambda_{\mathrm{sem}}=15$ and $\lambda_{\mathrm{align}}=1$ for both final variants, which provides a consistent operating point with a favorable balance between semantic retention and acoustic fidelity.

\end{document}